\documentclass{llncs}

\usepackage{xspace,float,amsmath,url}

\newcommand{\exclude}[1]{}

\begin{document}

\title{A note on the longest common Abelian factor problem}

\author{Szymon Grabowski}

\institute{
	  Lodz University of Technology, Institute of Applied Computer Science, \\
	  Al.\ Politechniki 11, 90--924 {\L}\'od\'z, Poland \\
	  \email{sgrabow@kis.p.lodz.pl}
}

\maketitle

\begin{abstract}
Abelian string matching problems are becoming an object of 
considerable interest in last years.
Very recently, Alatabbi et al.~\cite{AILR2015} presented the first solution 
for the longest common Abelian factor problem for a pair of strings, 
reaching $O(\sigma n^2)$ time with $O(\sigma n \log n)$ bits of space, 
where $n$ is the length of the strings and $\sigma$ is the alphabet size.
In this note we show how the time complexity can be preserved while 
the space is reduced by a factor of $\sigma$, and then how the time 
complexity can be improved, if the alphabet is not too small, 
when superlinear space is allowed.
\end{abstract}

\section{Introduction}
\noindent 
The longest common Abelian factor (LCAF) problem, posed at the 
String Masters 2013 meeting by Thierry Lecroq and Arnaud Lefebvre, 
can be stated like that:
Given two strings $A$ and $B$, both of length $n$, over the alphabet 
$\Sigma$, 
compute the maximal length of a factor in $A$ such that there exists 
a factor in $B$ being its permutation (i.e., being an Abelian match).
Moreover, it is desirable to return some (or all) occurrences of 
such factors in $A$ and $B$. 

To our knowledge, the only work on this problem was presented 
very recently by Alatabbi et al.~\cite{AILR2015}, in which 
they obtained $O(\sigma n^2)$ worst-case time 
with $O(\sigma n \log n)$ bits of space, 
where $n$ is the length of the strings and $\sigma$ is the alphabet size.
Further on, we will express the space in words, and the cited space 
becomes $O(\sigma n)$ words.

While the Alatabbi et al. algorithm is simple, 
let us note that the same result can be
immediately obtained by a reduction from a well-known problem, 
the (standard) longest common factor (LCF)\footnote{Also known 
as the longest common substring (LCS) problem. 
We prefer the word ``factor'' in the problem name, to avoid confusion 
with the abbreviation for the longest common subsequence.}.
Hui~\cite{H1992} showed that using a generalized suffix tree it is possible 
to find the LCF for a pair of strings of length $n$ in $O(n)$ time.
We use this algorithm $n$ times, for each factor length $\ell$, 
replacing each $\ell$ symbol long factor by its Parikh vector 
followed with a unique terminator (e.g., for the factors taken from $A$ 
the subsequent terminators can be 
$-1$, $-2$, ..., 
while 
for the factors taken from $B$ they can be 
$-n-1$, $-n-2$, ...).
The terminators disallow to have 
matches 
longer than $\sigma$.
If the found LCF is of length exactly $\sigma$, 
it must correspond to a pair of factors, one from $A$ and one from $B$, 
of length $\ell$.
This is obtained in $O(\sigma n)$ time for one value of $\ell$, 
using $O(\sigma n)$ space, hence the total time, 
for all possible factor lengths, becomes $O(\sigma n^2)$ with 
$O(\sigma n)$ space 
(we build and discard the generalized suffix trees one by one).
In this way, we obtained the same time and space as Alatabbi et al. did.

\section{Preliminaries}
Let $S$ be a string of length $n$ over an alphabet $\Sigma$ of 
size $\sigma = |\Sigma|$.
It can also be written as $S[1 \ldots n]$, where $S[i]$, $1 \leq i \leq n$, 
denotes its $i$-th symbol.
An analogous notation will be used for arrays.

Throughout the note we assume that $\sigma = O(n)$ and 
$\Sigma = \{1, 2, \ldots, \sigma\}$.
(If this is not the case, we can remap the alphabet for both 
input strings at the start with standard means, 
in $O(n\log n)$ time and $O(n)$ extra space.)

The Parikh vector for string $S$, 
denoted as $P(S)[1 \ldots \sigma]$, 
is defined as a vector (array) of size $\sigma$ storing the number 
of occurrences of each alphabet symbol in $S$.
Formally, $P(S)[c] = k$ iff $|\{i: S[i] = c\}| = k$, for any alphabet 
symbol $c$.
For two strings $S$ and $T$ of equal length and over a common alphabet,
we say that the Parikh vector $P(S)$ is (lexicographically) smaller 
than the Parikh vector $P(T)$, denoted as $P(S) < P(T)$, 
iff there exists an alphabet symbol $c'$, $1 \leq c' \leq \sigma$, such that 
$P(S)[c] = P(T)[c]$ for all $c < c'$ and $P(S)[c'] > P(T)[c']$.
The two Parikh vectors are equal, i.e., $P(S) = P(T)$, 
when $P(S)[c] = P(T)[c]$ for all symbols $c$.

\section{Reducing the space}
\noindent
First, let us note that recently Kociumaka et al.~\cite{KSV2014}
showed that for any tradeoff parameter $1 \leq \tau \leq n$, 
the LCF problem can be solved in $O(\tau)$ space and $O(n^2/\tau)$ time.
Applying this to the LCAF problem, we obtain 
$O(\tau \sigma n^2)$ time using $O(\sigma n/\tau)$ space, 
for any $1 \leq \tau \leq \sigma n$.

Yet, the specifics of LCAF allow for a better result.
We consider each factor length $\ell$ separately.
For a given $\ell$, we sort all $n-\ell+1$ factors of $A$ 
according to their Parikh vectors, using the LSD radix sort.
Each factor is represented as its start position in $A$.
There are $\sigma$ passes of the radix sort and accessing the keys' 
``digits'' seems to be the soft spot of this variant.
Yet, before each pass of the radix sort 
we scan $A$ and for each $\ell$-sized window collect the count 
of the corresponding symbol in it.
More precisely, just before the $i$-th pass of the radix sort, 
in which the keys will be distributed according to $P(\cdot)[\sigma-i+1]$, 
we compute and store $P(A[j \ldots j+\ell-1])[\sigma-i+1]$ for each factor $A[j \ldots j+\ell-1]$, 
using $O(n)$ time and $O(n)$ extra space.
Thanks to it, we can access a digit in the radix sort in constant time.
After the $i$-th pass, the $P(\cdot)[\sigma-i+1]$ statistics are discarded.
In this way, sorting of the $\ell$-long factors of $A$ takes $O(\sigma n)$ 
time and its output (and working area) requires $O(n)$ words of space.

We sort the factors of $B$ in the same way.
Additionally, for every $\sigma$-th evenly sampled $\ell$-long factor 
of $A$ and $B$, we store explicitly its Parikh vector using $O(\sigma)$ space.
More precisely, we compute and store the Parikh vectors for 
the factors $A[1 \ldots \ell], A[\sigma+1 \ldots \sigma+\ell], 
A[2\sigma+1 \ldots 2\sigma+\ell], \ldots$, 
and similarly for 
$B[1 \ldots \ell], B[\sigma+1 \ldots \sigma+\ell], 
B[2\sigma+1 \ldots 2\sigma+\ell], \ldots$.
As we scan the strings from left to right and compute the successive 
Parikh vectors incrementally (first making a copy of the previous vector), 
this phase takes $O(n + (n/\sigma)\sigma) = O(n)$ time and $O(n)$ space. 

The computed Parikh vectors serve to speed up factor comparisons 
during the last phase, which is to intersect the lists of factors 
from $A$ and $B$, similarly as in a binary merge operation.
Thanks to the Parikh vectors kept in regular intervals of $A$ and $B$, 
each factor comparison takes $O(\sigma)$ time, therefore the 
intersection takes $O(\sigma n)$ time.

The total cost of the described procedure, over all relevant factor lengths, 
becomes $O(\sigma n^2)$ and the required space is $O(n)$.
This matches the time complexity of the Alatabbi et al. solution, 
yet the space usage is decreased by a factor of $\sigma$.

\section{Reducing the time}

\subsection{The general variant}
\noindent 
In this section we present a variant which achieves $o(\sigma n^2)$ time 
for the price of superlinear space.
The key idea is to sort together factors of varying (yet close) lengths.

The whole sorting phase runs in $\Theta(n/k)$ steps, $k < \sigma$, 
where in the $i$-th step the factors of both $A$ and $B$ 
of all lengths from $ik+1$ to $(i+1)k$ 
are considered (yet, each group of factors, defined by their length, 
is sorted separately).
The required space grows to $O(kn)$.
To improve the time complexity, it is crucial to perform one step 
in $o(k\sigma n)$ time.
To this end, we make use of a data-oblivious sorting algorithm. 
An algorithm is called data-oblivious if its sequence of 
possible memory accesses is independent of its input values.
There exist such sort algorithms working in $O(n\log n)$ worst-case 
time (assuming that keys can be accessed in constant time), 
see~\cite{G2014} and references therein.

In our scenario, we compare the Parikh vectors of two factors 
of length $ik+1$ in $O(\sigma)$ time 
and also collect all the positions $i$, $1 \leq i \leq \sigma$, 
at which the respective Parikh vectors have different values.
These positions are inserted in bulk into a balanced binary search 
tree $\mathcal{T}$, in $O(\sigma)$ time.
Let the two factors be $A[u \ldots u+ik]$ and $A[v \ldots v+ik]$.
The next comparison concerns the factors of length $ik+2$: 
$A[u \ldots u+ik+1]$ and $A[v \ldots v+ik+1]$.
Their Parikh vectors can be obtained with updating only one counter 
in the previous vectors, which can also affect $\mathcal{T}$, 
as up to two elements should now be added to $\mathcal{T}$ 
and up to two elements should be removed $\mathcal{T}$. 
The operations on $\mathcal{T}$, including finding its minimum 
(or finding out that $\mathcal{T}$ is empty), 
which immediately serves to resolve the factor comparison, 
take $O(\log|\mathcal{T}|) = O(\log\sigma)$ time.
Similarly we handle the next pairs of factors, up to length $(i+1)k$.
Each time when equal (in the Abelian sense) factors are found 
and one of them is from $A$ and the other from $B$, 
we record their starting positions (in $A$ or $B$) and length.
In this way, we cannot miss the longest Abelian matching factors.
Note that in a comparison based sort, 
and in particular in a deterministic data-oblivious sort,
it is impossible not to compare 
equal items at some moment, if such exist.
To see this, imagine that we associate a real number with 
each item according to the sorted order; that is, the smallest item 
will have the smallest number and the largest item the largest number, 
and equal items will have equal associated numbers.
Now, if two items, $x$ and $y$, are equal and no other item in the collection 
is equal to $x$, not comparing $x$ to $y$ in the sorting process 
would mean that $x$ and $y$ are indistinguishable.
If, say, after the sorting $x$ stands (just) before $y$ 
and imagine $x$ is modified in such a way that its associated value gets 
greater by $\varepsilon/2$, where 
$\varepsilon$ is the minimum absolute difference between the associated values 
for any non-equal items in the collection, the hypothetical
sort algorithm not comparing $x$ to $y$
would produce the same output as before, which of course means 
that the algorithm is incorrect.

One step of the presented sort algorithm takes $O((\sigma + k \log\sigma)n\log n)$ time, 
which sums up to $O( (\sigma/k + \log\sigma)n^2 \log n)$ time over all steps, 
and the space usage is $O(kn)$.
Note that a space-time tradeoff is obtained with $k$ between 
$2$ and $\sigma/\log\sigma$.
For example, we can set $k = \sqrt{\sigma}$, which gives 
$O(\sqrt{\sigma} n^2 \log n)$ time and $(\sqrt{\sigma}n)$ space.
This time complexity is $o(\sigma n^2)$ when $\sigma = \omega(\log^2 n)$.


\subsection{Faster, sometimes}
\noindent 
In the algorithm above, the Parikh vectors of factors 
of length $ik+1$ were compared in $O(\sigma)$ time.
Let us try to reduce this time, trying to obtain a better 
overall space-time tradeoff.

To this end, for each length $ik+1$ 
we compute and store the Parikh vectors for 
factors of $A$ and $B$ sampled every $d$-th position, 
where $d < \sigma$ will be chosen later.
Additionally, we compute the positions of the differences between each of the 
$\Theta(n^2/d^2)$ pairs of Parikh vectors, storing them in a 
balanced binary search tree, as described in the previous subsection.
This requires overall $O(\sigma n^3/(d^2 k))$ extra time 
and $O(\sigma n^2/d^2)$ extra space.
However, the ``main'' time component gets reduced to 
$O( (d/k + \log\sigma)n^2 \log n)$.
As we are interested in improving the space-time tradeoff, 
we need to check if $d$ can be set to such value that 
the space complexity is not compromised, yet the time complexity 
improves, at least for some $k$ and $\sigma$.
Clearly, it requires that $\sigma n^2/d^2 = O(kn)$, i.e., 
$d = \Omega(\sqrt{\sigma n/k})$.
As only $d = o(\sigma)$ may improve the time complexity, 
we need to have $n/k = o(\sigma)$ (and of course $\sigma = \omega(1)$).
An extra requirement is $k = o(\sigma/\log\sigma)$.
Finally, improving the time complexity means that 
$(d/k + \log\sigma)n^2 \log n + \sigma n^3/(d^2 k) =
o((\sigma/k + \log\sigma)n^2 \log n)$, which does not introduce 
an extra constaint since $\sigma n^3/(d^2 k) = O(n^2)$, given 
the aforementioned lower bound on $d$.

We set $d = \Theta(\sqrt{\sigma n/k})$.
This implies $d = \omega(\sqrt{n\log\sigma})$ and thus also 
$\sigma = \omega(\sqrt{n\log n})$, which eventually gives 
$d = \omega(\sqrt{n\log n})$.

To sum up, if $\sigma = \omega(\sqrt{n\log n})$ and 
$k = \omega(n / \sigma)$ but also $k = o(\sigma/\log\sigma)$,
by choosing $d = \Theta(\sqrt{\sigma n/k})$
we preserve the $O(kn)$ space and improve the time 
to $O((\sqrt{\sigma n/k^3} + \log\sigma)n^2 \log n)$.
In most cases the improvement is not large: 
for example, if $\sigma = n^{0.8}$ and $k = n^{0.4}$, 
the time complexity is slashed by a factor of $n^{0.1}$.
On the other hand, if e.g. $\sigma = \Theta(n/\log n)$ 
and $k = \Theta(n^{2/3}/\log n)$, then the time complexity 
becomes $O((\log\sigma) n^2 \log n)$, 
an improvement by a factor of $n^{1/3}$.

\section{Conclusions}
\noindent 
Finding the longest common Abelian factor is a recently posed problem, 
with a solution given in~\cite{AILR2015}, 
achieving $O(\sigma n^2)$ worst-case time and 
needing $O(\sigma n)$ words of space.
A significant weakness of that result is its space requirement, 
which may be unacceptable with a larger alphabet.
In this work we improve this result in two ways.

One algorithm keeps the time complexity of the previous result, 
while it reduces its space to $O(n)$.
This is obtained with very simple means (the key component is 
the LSD radix sort).
The other algorithm of ours increases the space to $O(kn)$
and achieves the time complexity of $O( (\sigma/k + \log\sigma)n^2 \log n)$, 
where $k \leq \sigma/\log\sigma$ is a freely chosen parameter.
When $\sigma = \omega(\log n \log\log n)$ it is always possible to 
choose such $k$ that this algorithm beats the result from~\cite{AILR2015} 
in both time and space complexity.
This variant is also simple conceptually, yet it makes use 
of a deterministic data-oblivious sort algorithm of optimal complexity 
in the comparison based model.
There are several such algorithms known, but none of them is really simple. 
A more practical choice could be the textbook Shell sort algorithm 
with the sequence of gaps of the form $2^p 3^q$, proposed 
by Pratt in 1972~\cite{P1972}.
Applying this Shell sort variant would deteriorate our time complexity 
by a factor of $\log n$.
The latter of the two algorithms is also improved slightly 
for convenient values of $\sigma$ and $k$.

We are convinced that better algorithms for the LCAF problem are possible.
One obvious line of attack is using word-level parallelism
(in the word-RAM model) for Parikh vector comparisons.
The anticipated speed-up factor is however only about $w / \log(n/\sigma)$, 
where $w$ is the machine word size.
A more interesting question is whether sharing computations 
for different factor lengths could be exploited with a stronger effect 
than presented here.

\bibliographystyle{abbrv}
\bibliography{lcaf}

\end{document}